\documentstyle[prc,aps,epsfig,twocolumn]{revtex}

\begin{document}

\twocolumn[\hsize\textwidth\columnwidth\hsize
           \csname @twocolumnfalse\endcsname
\date{\today}
\title{{\bf From microscopic to macroscopic dynamics in mean-field theory:
effect of neutron skin on fusion barrier and dissipation.}}
\author{Denis Lacroix}
\address{{\it LPC/ISMRA, Blvd du Mar\'{e}chal Juin, 14050 Caen, 
France}\\
\begin{abstract}
In this work, we introduce a new method to reduce the microscopic 
mean-field theory to a classical macroscopic dynamics during the initial 
stage of
fusion reactions. We show that TDHF (Time-dependent Hartree-Fock) could be a
useful tool to gain information on fusion barriers as well as on one-body
dissipation effects. We apply the mean-field theory to the case of 
head-on reaction between $^{16}$O and $^{16,22,24,28}$O in order 
to quantify the effect of neutron skin on fusion. 
We show that the determination of fusion barrier requires, in addition
to a precise knowledge of the relative distance between the 
center of mass of the two fusing nuclei, the
introduction of an additional collective coordinate that 
explicitly breaks the neutron-proton symmetry. 
In this context, we estimate the position, height 
and diffuseness of the barrier as well as the one-body friction
and show that a global enhancement of the fusion cross-section is expected
in neutron rich nuclei.
\end{abstract}
}
\maketitle


\vskip2pc
]

\section{introduction}

The understanding of the formation of compact system during fusion 
of two nuclei implies the development of nuclear models 
which can take into account quantum effects, the dynamical evolution and 
dissipation in a common framework. From that point of view, 
the description of the conversion of
relative motion into internal excitation of the composite system and also the
deexcitation of the compound nucleus is far from being complete. 
Due to
the diversity of dissipative mechanisms, the microscopic descriptions 
of fusion
reactions is often substituted by a macroscopic dynamics in which only few relevant
degrees of freedom are considered. 
Such macroscopic description are only valid under a series of assumption 
like separation in time-scales between relevant and other degrees of freedom.
The validity of these hypothesis as well as the link with microscopic 
theories is under debate.
 
In the present work, starting from quantum
mean-field theory, we address the problem of the reduction 
of the microscopic dynamics to a macroscopic theory\cite{Nor80,Sch84,Fro96,Abe95,Hof97}. 

\subsection{Mean-field theories}

\noindent
We consider the Time-Dependent  
Hartree-Fock (TDHF) theory: 
one-body degrees of freedom are supposed to be dominant 
and the description of the system reduces to the time evolution of 
a Slater-determinant $\left| \Psi \right>$ in a self-consistent mean-field.
Noting $\rho$ the associated one-body density matrix, the evolution is given by
the Liouville equation\cite{Rin80} 

\begin{eqnarray}
i\hbar \frac{d\rho}{dt} = \left[h[\rho], \rho \right]
\end{eqnarray}
where $h\left[\rho\right]$ is the self-consistent field.
Such a picture has been widely used to investigate nuclear dissipative 
reactions, and gives reasonable results in the case of fusion 
reactions\cite{Bon76,Neg82,Dav84}. In addition, it has the advantage 
to take into account shell effects and to include the so-called one-body 
dissipation.

However, fusion reactions and more generally deep inelastic 
reactions are often described assuming that some macroscopic variables
are dominant\cite{Sch84,Fro98}. Although a large number of work 
is devoted to this subject, the link between macroscopic dynamics and
microscopic approaches such as TDHF is still unclear\cite{Wei80}.

In this paper, we address this problem considering head-on 
reactions. In this case, the relative distance $R$ between nuclei in the
entrance channel is known to describe the main features of the dynamics
and its evolution follows the classical Hamilton-like equation
\footnote{For
simplicity, we have written the expression in its simplest form, it should
however be noted that other degrees of freedom like angular 
momentum\cite{Fro96} or mass
assymetry\cite{Gia00} may be considered in the formalism.}\cite{Fro96}: 

\begin{eqnarray}
\left\{
\begin{array} {ll}
\frac{dR}{dt} =& P / {\mu(R)} \\
&  \\
\frac{dP}{dt} =& F(R) - \gamma \left(R\right) \dot{R}   \\
\end{array}
\right.
\label{clas}
\end{eqnarray}
where $\mu(R)$ is the reduced mass of the system 
and the force is assumed to derive from a potential (
$F(R)=-\frac{\partial V}{\partial R}$). 
The dissipation kernel $\gamma(R)$ 
is supposed to reproduce the effect of the other non described 
degrees of freedom on the considered variables.
 
The dissipative effects  are generally  taken
by including the one-body dissipation 
in a liquid drop picture\cite{Swi84}
or using an adiabatic approximation in a two-center shell model 
(see for instance \cite{Mos80}). 
However, the understanding of dissipative effects in
nuclear collisions is far from being complete and the use 
of equation (\ref{clas}) rises some questions. A large number 
of effects are known to participate to the dissipation of 
the relative to the internal excitation energy during the reaction. 
Among them, exchange of particles and deformation are 
generally assumed to be dominant and are accounted as
one-body dissipation effects.
We would like however 
to mention that the dissipative process described by
equation (\ref{clas}) is obtained using quantum 
perturbation theory \cite{Gro75}. 
It is not clear whether it can be applied to the case of 
violent collisions like fusion. 

Mean-field theory like TDHF does a priori includes one-body 
dissipation in a non-perturbative framework. In addition, it
assume neither separation between degrees of freedom 
nor adiabaticity. However, the link with the so-called 
one-body dissipation
theory is still under discussion\cite{Bau95}. 

It should also be noted that friction models assumes 
a fast equilibration of other degrees of freedom as compared 
to the relative motion leading to the formation of a compound nucleus 
at finite temperature. Due to the absence of two-body or higher 
order correlations, TDHF is unable to account for this effect. 
From the experimental point of view, the equilibration time
is estimated to be of the order of the reaction time \cite{Bor92}, 
which means that the classical description assuming
thermal equilibration assumption is questionable as far as 
fusion is concerned.

In this paper, we discuss the possibility to use macroscopic
equations for TDHF although it does not assume perturbation 
theory or equilibration. In the following, we show that such 
a reduction seems possible and gives reasonable results on fusion 
barriers properties.





Different works already exist on 
the possibility of such a reduction 
of microscopic theory 
\cite{Koo79,Bri81,Jor87,Pal88}. 
However, as it was already noted in \cite{Koo79}, high 
numerical accuracy as well as 
the use of complete realistic effective forces is required.
In this article, we use the three-dimensional TDHF code developed by Bonche and
co-workers \cite{Kim97} and the 
full SLy4d Skyrme force including spin-orbit \cite{Cha95}. 

In the following, we apply the mean-field theory to head-on reactions 
between oxygen isotopes. The static properties of oxygen 
isotopes are first presented. We then describe the method used 
to reduce mean-field evolution 
to the relative distance only and extract quantitative 
information on the fusion process.
The macroscopic potential landscape and the dissipative aspects 
are then discussed.

\section{Results}

\subsection{Structure of oxygen isotopes}

\begin{figure}
\begin{center}
\epsfig{file=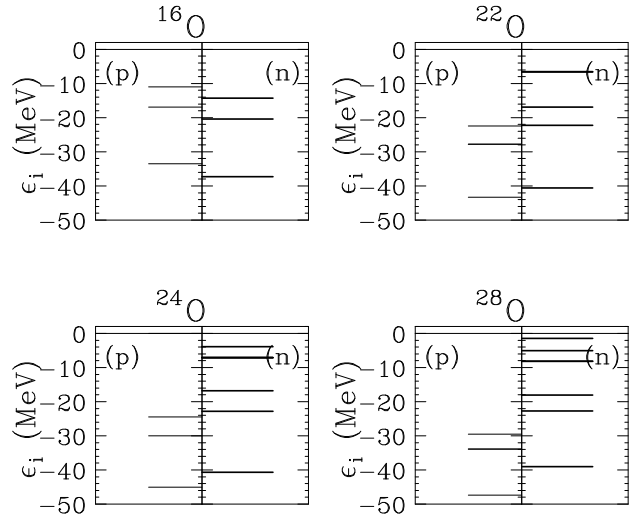,height=7.cm}
\end{center}
\caption{Proton (p) and neutrons (n) energy levels obtained in 
Hartree-Fock calculation described in the text using the Sly4d 
effective Skyrme force for 
$^{16}$O, $^{22}$O, $^{24}$O, $^{28}$O isotopes.}
\label{fig:level}
\end{figure}

For each of the considered nuclei, the self-consistent Hartree-Fock 
equation reads:
\begin{eqnarray}
h[\rho] \left|\varphi^\tau_i \right> = 
\varepsilon_i \left|\varphi^\tau_i \right>, 
\label{hf}
\end{eqnarray}
where $h[\rho]$ is the mean-field hamiltonian,
$\left|\varphi^\tau_i \right>$ denotes the single particle wave functions 
and $\tau$ denotes the isospin quantum number 
($\tau$ = p [proton] or $\tau$ = n [neutron]).
Equation \ref{hf} 
is solved in r-space on a mesh of size 
$\left(16 fm\right)^3$ with a step of $0.8$ fm. 
The complete expression of $h[\rho]$ for Skyrme forces can 
be found for instance in \cite{Cha98}. 
The resulting single-particle levels energies obtained using the 
SLy4d Skyrme force are presented in Fig. 
\ref{fig:level} for the four isotopes $^{16}$O, $^{22}$O, 
$^{24}$O and $^{28}$O.

\begin{figure}
\begin{center}
\epsfig{file=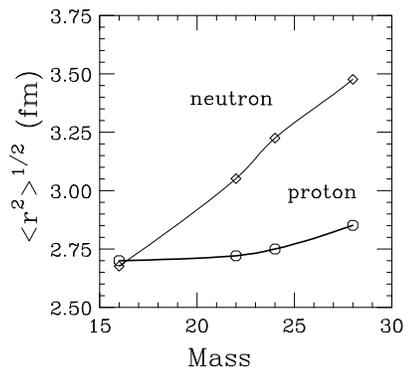,height=5.cm}
\end{center}
\caption{Proton (circle) and neutron (diamond) root mean square radius 
predicted by the 
Hartree-Fock calculation using the Sly4d Skyrme interaction for the 
$^{16}$O, $^{22}$O, $^{24}$O, $^{28}$O isotopes.}
\label{fig:rms}
\end{figure}

The associated neutron and 
proton root mean square radius  (RMS) are reported in
figure \ref{fig:rms}. These root mean square radii leads to 
RMS matter radii in agreement with experimental data for the 
$^{16}$O, $^{22}$O and $^{24}$O nuclei \cite{Oza01}.
A neutron skin 
is evidenced when the number of 
neutrons increases.
 
Once the static properties are obtained, we consider the evolution of 
the two colliding nuclei in a larger volume $\left(32 fm\right)^3$ with the 
same step in r-space as before. The initial distance between the 
center of the two
nuclei is equal to 16 fm. The time step is chosen as $0.45$ fm/c. 
Before showing the results of the dynamical calculation, we discuss 
the method used to reduce the mean-field dynamics to the time 
evolution of a few macroscopic coordinates.

\subsection{Reduction to macroscopic dynamics}

In order to define the relative distance between the two nuclei, 
we consider the 
mean-field equation in r-space and the associated one-body density 
$\rho^{\tau}(r,r')$ with
$\rho^\tau = 
\sum_{i} \left|\varphi^\tau_i(t) \right>\left<\varphi^\tau_i(t)\right|$. 
We assume, as in ref. \cite{Bon84}, 
that at each time step, the system could be split 
into two subspaces 1 and 2 with the associated densities 
$\rho^\tau_x$ ($x=1,2$). 
We define the separation plane after the contact at the neck 
position as it is done generally in macroscopic models\cite{Has88}.


Once the separation is obtained, we compute for each subspace, the 
number of neutrons $N_x = Tr\left(\rho^n_x\right)$ or protons 
$Z_x = Tr\left(\rho^n_x\right)$,
the mean position 
\begin{eqnarray}
R^\tau_x \equiv <\hat{r}>_x = Tr\left(\hat{r} \rho^\tau_x \right)/
Tr\left(\rho^\tau_x\right)
\end{eqnarray}
and the associated momentum 
\begin{eqnarray}
P^\tau_x \equiv <\hat{p}>_x = Tr\left(\hat{p} \rho^\tau_x \right)
\end{eqnarray}
where $Tr$ denotes the Trace. 
Let us first assume that we can neglect the isospin degree of freedom 
(the validity of this assumption will be discussed in the following).
We define the center of mass as well as the associated total 
momentum: 
\begin{eqnarray}
R_x  &=& \left( N_x  R^n_x + Z_x  R^p_x \right)/A_x \\
P_x &=& P^n_x + P^p_x.
\end{eqnarray}
where we have introduced $A_x = N_x + Z_x$. \\
Using Hamilton equation for each subsystem, we estimate 
the mass from equation \ref{clas}
\begin{eqnarray}
m_x = P_x/\left(\frac{dR_x}{dt}\right)
\end{eqnarray}
At a given relative distance $R = R_1 - R_2$, we compute
the reduced mass associated with the composite system
$\mu(R) =  m_1 m_2 / (m_1+m_2)$. Note that, since we have 
a Skyrme force, with in particular $t_1$, $t_2$ and spin-orbit terms, 
$\mu(R)$ is different from the normal value:
\begin{eqnarray}
\mu_0(R) = m\frac{A_1(R)A_2(R)}{(A_1(R)+ A_2(R))} 
\end{eqnarray}
where $m$ is the nucleon mass. 
In our calculation, we obtain $\mu(R)/\mu_0(R) \simeq cte \simeq 0.67$ 
for all systems. This remains valid up to the formation of a compact shape.

Once these quantities have been defined and calculated, 
the different components entering in Eq. (\ref{clas}): $R(t)$, $dR(t)/dt$,
$P(t)$ and its derivative can be
computed.  

Eq. \ref{clas} contains two unknown quantities, the 
force $F(R)$ and the friction coefficient $\gamma (R)$. In order 
to estimate these quantities, a method proposed in ref. 
\cite{Koo77,Koo79} is used. It consists in performing mean-field evolution
with two slightly different initial energies $E^{I}$, $E^{II}$.
 At a given position $R$, Eq. \ref{clas} can be interpreted as a system 
of two equations with two unknown quantities which reads:
\begin{eqnarray}
\left(
\begin{array}{cc}
1 &  -\frac{d R^I}{dt}   \\
1 &  -\frac{d R^{II}}{dt} \\
\end{array}
\right) 
\left(
\begin{array}{c}
F\left( R \right) \\
\gamma (R)
\end{array} 
\right)
=
\left(
\begin{array}{c}
 \frac{dP^{I}}{dt} \\
 \frac{dP^{II}}{dt} \\
\end{array} 
\right)
\label{eq:inv} 
\end{eqnarray}
where all derivatives are taken at position R.
We checked the numerical stability of the proposed procedure, by solving the classical 
equation associated with the same initial condition as TDHF evolution and 
with the friction and potential extracted from it. As expected, we recover 
the TDHF evolution.

\subsection{Nuclear potential}

\begin{figure}
\begin{center}
\epsfig{file=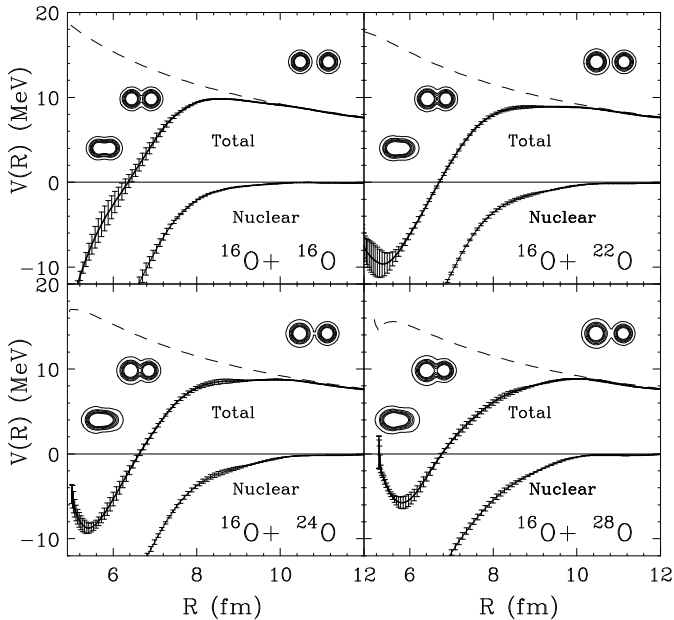,height=9.cm}
\end{center}
\caption{Potential landscape obtained in head-on 
$^{16}$O+$^{16,22,24,28}$O reactions. 
The total (thick solid lines), coulomb (dashed line) and
nuclear potential (thin solid lines) are presented. 
For the calculation, different couples
of trajectories taken within the initial energies $E/A = 0.75, 0.70, 0.65, 0.60$
MeV are presented. "Errorbars" correspond to the 
mean-variance on results 
obtained with different initial conditions. 
The isodensity plots in the reaction plane 
are also superimposed for different times.}
\label{fig:pot}
\end{figure}

We performed TDHF calculations with four different initial center of mass
energies $E/A = 0.60, 0.65, 0.70$ and $0.75$ MeV for each reaction.
Figure \ref{fig:pot} gives the potential for the reactions 
$^{16}$O+$^{16,22,24,28}$O obtained by inversion of Eq. \ref{eq:inv}. 
To extract the nuclear part, we assume  
$V_{nuclear}(R) = V(R) - V_C(R)$, where the coulomb part is simply taken as 
$V_C(R) = e^2 Z_1(R) Z_2(R) / R$ (dashed line). 
For each reaction, we have solved Eq. \ref{eq:inv} using different couple 
of TDHF evolution with different initial energies. The solution depends slightly 
on the couple of trajectories chosen, this uncertainty is represented by
errorbars in Fig. \ref{fig:pot}. 
The uncertainties are rather small indicating that the 
microscopic dynamics is already 
well described by Eq. \ref{clas}. 

Examining the global shape of $V(R)$, and in particular 
the maximum, we see that although the barrier height is slightly
affected by the neutron skin, the increase of the neutron number not only
increases the barrier position $R_B$ but also change its 
diffuseness. This effect should normally indicate a very 
different behavior in the sub-barrier fusion of very neutron rich nuclei as
compared to stable isotopes. Using a different technique, the same
conclusion is drawn in ref. 
\cite{Chr95} while in ref.
\cite{Kim94,Kim97}, it is concluded that the excess of 
neutron does not affect fusion. This point is discussed in the 
following

\subsection{Dissipation}

\begin{figure}
\begin{center}
\epsfig{file=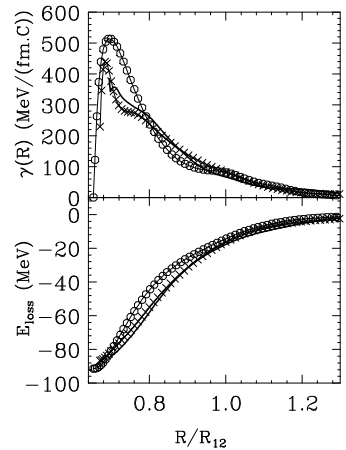,height=6.cm}
\end{center}
\caption{Top: Friction coefficient as a function of relative distance 
of head-on $^{16}$O+$^{16}$O (circles), $^{16}$O+$^{22}$O 
(crosses) and $^{16}$O+$^{28}$O (thick lines) reactions. 
Bottom: Associated integrated Rayleigh function $E_{loss}$ 
as a function of the relative distance normalized to the sum of 
the nuclear radius of the target and projectile $R_{12} = \sqrt{\frac{5}{3}}
\left(<r^2_T>^{1/2} + <r^2_P>^{1/2}\right)$. In these figures errorbars 
are not shown. For simplicity, we do not show the $^{16}$O+$^{24}$O which is
similar to the $^{16}$O+$^{22}$O case.}
\label{fig:gam}
\end{figure}

While numerous works have been devoted to potentials determination, 
only few quantitative analyses exist on the dissipation effects 
starting from a non-adiabatic microscopic theory. The method described above
gives also information on the dissipative processes.  
The friction coefficient associated to the mean-field evolution 
of the system can be computed. It is  
presented in Fig. \ref{fig:gam}.

In order to have a global information on the amount of energy transferred
from the relative motion into internal degrees of freedom of the compound 
system, the integrated value 
of the Rayleigh function can be evaluated. 
The Rayleigh function gives the rate of energy
dissipated by unit of time: 
\begin{eqnarray}
E_{ray}(t) = dE/dt = -\gamma(R) \dot{R}^2
\end{eqnarray}
and its integrated value $E_{loss}(R(t))$ reads (Fig. \ref{fig:gam}):  
\begin{eqnarray}
E_{loss} (t) = 2\int_{t_0}^{t} E_{ray} (s) ds 
\end{eqnarray}
In order to compare different systems, $\gamma(R)$ and $E_{loss}(R(t))$
are plotted as a function of the relative distance normalized to 
$R_{12} = \frac{5}{3}\left(<r^2_T>^{1/2} + <r^2_P>^{1/2}\right)$ where 
$P$ and $T$ refers respectively to the projectile and target. It is worth
noticing that dissipation is larger at larger distances than 
$R/R_{12} \sim 0.7$ in the case of neutron rich nuclei.
However, at intermediate distance $0.5<R/R_{12}<0.7$ the  
dissipation is larger in the symmetric case.
We indeed expect a different behavior 
between symmetric and assymetric systems. If the one-body
dissipation picture is correct, part of the viscosity is due to the
change of the global shape of the system (the so-called
wall dissipation \cite{Blo78}). In assymetric reactions, this effect should 
be completed by the friction due to the
exchange of nucleons between the two subsystems (the window dissipation 
\cite{Ran80}). The link between the friction coefficient in TDHF and the 
wall plus window picture is however not clear. Indeed, mean-field 
theory also includes other degrees of freedom which are not taken into account
in the wall and window formula. In particular, a fraction of 
the initial energy is converted into
collective excitation (giant resonances). In our case, this effect is also
integrated in $\gamma(R)$.

We finally would like to mention that, although the dissipation is different 
between symmetric and assymetric systems, the value of $E_{loss}$ at minimal
distance of approach is independent of the assymetry.

\subsection{Precise determination of the barrier properties}

\begin{figure}
\begin{center}
\epsfig{file=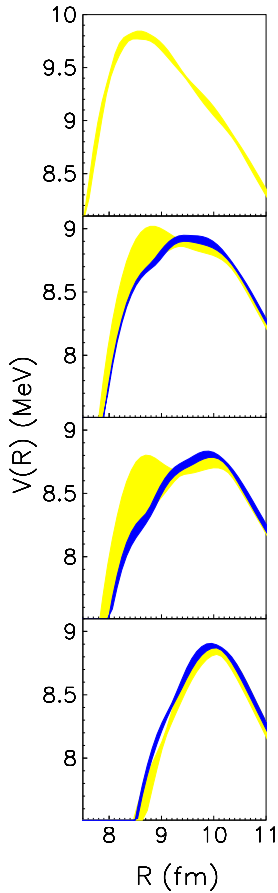,height=12.cm}
\end{center}
\caption{ Zoom of the fusion barrier region 
obtained for  
$^{16}$O+$^{16,22,24,28}$O reactions respectively from top to bottom
using the macroscopic 
interpretation of TDHF. 
Filled areas corresponds to calculated potential with
uncertainties, the gray area corresponds to the result obtained using
Eq. \ref{clas} while the black area is the result obtained including 
the isospin relative distance
difference (see text). 
In the case of $^{16}$O+$^{16}$O there is a simple curve since 
the additional degrees of freedom 
does not change the extracted potential.}
\label{fig:potzoom}
\end{figure}

The possibility to obtain fusion cross-section from mean-field 
requires a precise estimate of the nuclear potential in the vicinity of the
fusion barrier.
Details of the fusion barrier extracted using Eq. \ref{clas} are shown 
in figure \ref{fig:potzoom} (gray area). 
For $^{16}$O+$^{22}$O and 
$^{16}$O+$^{24}$O reactions, rather large uncertainties are observed for
specific relative distances. 
These uncertainties represents the 
dependence of the potential $V(R)$ with the initial energies, 
i.e. the polarisation of the 
extracted solution using Eq. \ref{clas} with mean-field trajectories. 
In our microscopic calculations, this polarisation could 
be directly attributed to the appearance of 
another important degree of freedom
which is not included in Eq. \ref{clas}. Indeed, regions where uncertainties
are important are associated with relative 
distances between protons in the two nuclei
differing from the distances between neutrons. In order to account for
this additional effect, we have introduced the difference between 
the proton and neutron relative distances:
\begin{eqnarray}
\Delta R_{np} = \left(R^n_1 - R^n_2\right) - \left(R^p_1 - R^p_2\right) 
\end{eqnarray}        
and assumed that the Eq. \ref{clas} is modified in order to include this
additional degree of freedom: 
\begin{eqnarray}
\frac{dP}{dt} &=& F(R) - \gamma \left(R\right) \dot{R} 
- \gamma_{np}(R) \Delta \dot{R}_{np}
\label{clas_np} 
\end{eqnarray}
As in the previous case, using three different initial conditions, 
Eq. \ref{clas_np} can be interpreted
as three linear equations with three unknown
quantities $F(R)$, $\gamma \left(R\right)$ and  $\gamma_{np}(R)$ (see
Eq. \ref{inver}). 
The inversion of such ensemble of equations is rather difficult since the
additional degree of freedom is not always important during the evolution.
This often leads to an overcomplete set of equations. 
A discussion on inversion method is given in appendix A. 

Assuming that the dynamical mean-field theory can be reduced to 
Eq. \ref{clas_np}, we obtain the new potentials represented 
by a black area in Fig. \ref{fig:potzoom}. In that case, the
initial energy dependence are strongly reduced as compared to the previous 
calculation indicating the importance of this new degree of freedom. 
Note that in the symmetric reaction $^{16}$O+$^{16}$O, the additional 
degree of freedom can be neglected, and the obtained potential strictly
corresponds to the old one.
It should be noticed that 
the second maximum present in  
$^{16}$O+$^{22}$O and $^{16}$O+$^{24}$O reactions using Eq. \ref{clas} 
has disappeared.
In order to quantify more precisely the effect of the neutron excess 
on fusion,
we have fitted the new value of $V(R)$ in the vicinity of 
the barrier assuming 
$V(R) = V_B - \Delta V_B (R -R_B)^2$. We have reported in table \ref{tab:vbrb},
the barrier height $V_B$, the peak position $R_B$ and the deduced curvature 
$\hbar \omega_B$ defined as $\hbar \omega_B = \hbar \sqrt{\Delta V_B/\mu_0}$.   

In the case of $^{16}$O + $^{16}$O, where experimental data exist, 
a good agreement with the extracted barrier height and
position \cite{Vaz81,Hai84} is found. For other reactions, experimental 
results are not available. We would like to mention that 
we have also applied this method to the formation of heavy and very 
heavy elements. In all cases, we found a good matching with the experimental fusion properties 
giving additional confidence in the physical meaning of the extracted 
barriers.  

Looking at the three assymetric reactions,  
the barrier height is almost constant while, as expected, the peak position 
increases slowly as well as the barrier diffusion. However these 
trends could not be extended to the symmetric case. This fact
associated to the fact that it is not necessary to introduce additional degrees
of freedom in the symmetric reaction case, shows the 
peculiarity of the neutron excess in fusion reactions.
 
\begin{figure}
\begin{center}
\epsfig{file=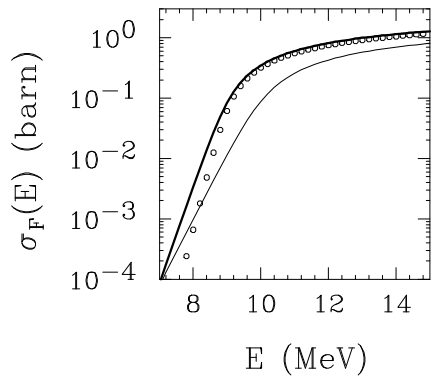,height=5.cm}
\end{center}
\caption{Example of calculated fusion cross-section using the Wong formula for
the reaction $^{16}$O+$^{16}$O (thin line), $^{16}$O+$^{22}$O (circles)
and $^{16}$O+$^{28}$O (thick line).}
\label{fig:sig}
\end{figure}

\begin{table} 
\begin{center}
\begin{tabular}{c|cccc|}
& $^{16}$O & $^{22}$O & $^{24}$O & $^{28}$O \\
\hline
$V_B$ (MeV) & 9.8 & 8.9 & 8.8 & 8.9 \\
$R_B$ (fm)  & 8.6 & 9.6 & 9.8 & 10.0 \\
$\hbar \omega_B$ & 2.39 & 1.21 & 1.36 & 1.65 \\
\hline
\end{tabular}
\end{center}
\caption{Fusion barrier $V_B$ and position extracted from head on 
collisions $^{16}$O+$^{16,22,24,28}$O.} 
\label{tab:vbrb}
\end{table}
In order to have some qualitative information on fusion cross sections, we
have injected the values reported in table \ref{tab:vbrb} in the Wong 
formula\cite{Won73}:
\begin{eqnarray}
\sigma_F(E) = \frac{\hbar \omega_B R_B^2} {2E} 
ln\left(1 + exp \left(2 \pi \left(E-V_B\right)/\hbar \omega_B\right)\right)
\end{eqnarray}
where E is the laboratory energy. The cross sections 
for $^{16}$O+$^{16}$O (thin line),
$^{16}$O+$^{22}$O (circles)
and $^{16}$O+$^{28}$O (thick line) reactions are displayed in Fig. 
\ref{fig:sig}.
For fusion above the barrier, due to the 1 MeV lower 
value of the barrier, a larger 
fusion cross section for the case of neutron rich isotopes as compared to
symmetric $^{16}$O reaction is expected.
However, there is no differences between $^{22}$O, $^{24}$O or  $^{28}$O
above $E>10MeV$. 

The excess of neutron is also predicted to modify the sub-barrier
cross-section \cite{Das92,Tak91,Sid01}: an increase of the sub-barrier 
cross-section
associated with a decrease of the slope of 
the cross-section is expected. 
In Fig. \ref{fig:sig}, a more complex dependence
in the slope of the cross-section is observed. It should however be noted 
that the inclusion 
of friction, which is neglected in the Wong formula, may change the 
slope in the sub-barrier region. Since the friction changes considerably
between stable and neutron rich nuclei (see Fig. \ref{fig:gam}), 
a comparison of the different slopes is difficult.
In the case of the two reactions $^{16}O+^{22}O$ and $^{16}O+^{28}O$, 
friction coefficients have the same behavior as a function of the 
relative distance and the cross-sections may be compared. 
In this case, we indeed observe a 
smaller slope in the $^{28}O$ case.

  
\begin{figure}
\begin{center}
\epsfig{file=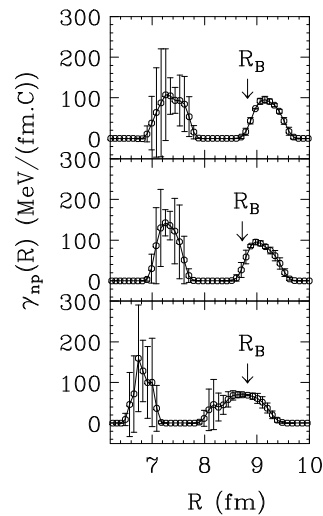,height=7.cm}
\end{center}
\caption{Top: Friction coefficient $\gamma_{np}$ 
as a function of the relative distance 
in head-on $^{16}$O+$^{22}$O (top), $^{16}$O+$^{24}$O (middle) 
and $^{16}$O+$^{28}$O (bottom) reactions. The barrier position 
is indicated by an arrow.}
\label{fig:gamnp}
\end{figure}

The evolution of $\gamma_{np} (R)$ as a function of the relative distance is
shown in Fig. \ref{fig:gamnp} as well as the
position of the barrier. The effect of the new coordinate is
intermittent. The first maximum of  $\gamma_{np}$ is close to the barrier. 
This is directly understood by the fact that, at this point, 
the difference 
between proton and neutron distances is maximum. Indeed, in this case, the
coulomb repulsion is large, which goes in favour of a larger proton distance
while, due to the existence of a neutron skin, the attractive nuclear 
part acts first on neutrons and thus tends to reduce the neutron distance.
After the contact, the effect of $\gamma_{np}$ appears periodically
due to the appearance of neutron-proton oscillations which are 
already present in the entrance 
channel.   

 The necessity to introduce explicitely the difference between 
neutron and proton center of mass in neutron-rich nuclei has already 
been noted in several work \cite{Sat91,Das92,Can95}. In all cases, 
it was concluded that the neutron excess is expected to increase the 
fusion cross-section as also observed in the present work. 
Different effects are indeed 
expected to act in this direction: the large extension of the neutron
skin which implies the influence of the nuclear field at larger 
relative distance, the coupling between the soft dipole mode 
with the relative motion. The time-dependent mean-field 
model includes all these effects. In particular, the 
observable $\Delta R_{np}$ is obviously connected to 
the polarisation of each nucleus before fusion. It should 
also be noted that $\Delta R_{np}$ is affected by the 
transfer of particles 
initiated just after the contact in order to equilibrate 
the $N/Z$ ratio between the two partners. This transfer
which is known to induce damping in the relative motion, gives
rise to the excitation of a pure collective GDR in the composite 
nucleus\cite{Sim01}.
 
Although the additional friction term in Eq. (14) seems to 
strongly reduce uncertainties on the extracted  potentials, for longer 
time, a more complex dynamics is expected. In particular, 
it would be convenient to treat also collective processes 
like the excitation of giant resonances. For instance, the 
effect of polarisation as well as the onset of the GDR after 
fusion requires the introduction of potential effect coming
from the symmetry energy.

To quantify the amount of energy transferred 
from the relative motion 
to the new coordinate, we have also computed a quantity similar
to $E_{loss}$ noted $E_{np}$ which is defined as:

\begin{eqnarray}
E_{np} (R) = -\int_{t_0}^{t} \gamma_{np}(R(s)) \dot{R}(s) \Delta \dot{R}_{np}(s)ds
\end{eqnarray} 

\begin{figure}
\begin{center}
\epsfig{file=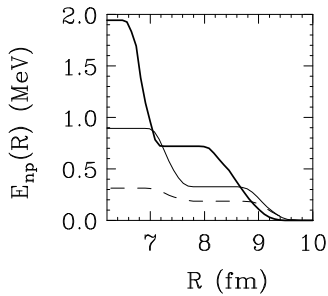,height=4.cm}
\end{center}
\caption{ Energy transferred between the relative motion and the neutron to 
proton oscillation as a function of the relative distance for the reaction
$^{16}$O+$^{22}$O (dashed line), $^{16}$O+$^{24}$O (thin line) 
and $^{16}$O+$^{28}$O (thick line).}
\label{fig:enp}
\end{figure}

The result of the computation is shown in Fig. \ref{fig:enp}. The 
amount of transferred energy to the new observable 
is directly proportional to the neutron excess.

\section{Conclusion}

In this paper, we have shown that the properties of a macroscopic dynamics 
can be rather precisely inferred from a microscopic 
quantum mean-field dynamics by reducing the one-body information to few
collective coordinates. 
In order to have quantitative information 
on the barrier properties, the "naive" 
reduction taking into account only the relative distance 
is not sufficient in neutron rich nuclei and it is necessary to introduce the 
difference between 
neutrons and protons during the onset of fusion. 

We have applied our method to the case of 
head-on oxygen reactions involving oxygen isotopes.  
We have shown that, while the fusion barrier is only slightly affected by the
neutron skin, the barrier shapes change. We thus expect that 
the neutron skin will modify preferentially the sub-barrier fusion reactions.

Beside the potential landscapes, the mean-field dynamics 
gives also information on how the relative motion dissipates into internal 
excitation of the composite system. Such a dissipation which gives
directly the associated friction coefficient has rarely been obtained in
non-adiabatic microscopic theory. We have shown that during 
the initial stage of the reaction, part of the dissipation corresponds 
to an isospin symmetric 
dissipation of the relative motion ($\gamma$). In this case, the total 
amount of energy deposited in the system seems independent of the neutron skin.
On the other side, another part of the dissipation associated with
the 
neutron-proton difference ($\gamma_{np}$) depends explicitely on 
the neutron excess.

In this work, we have shown how mean-field theory which includes nuclear structure
effects and does not invoke specific approximations on macroscopic dynamics,
may be very useful to extract quantitative information on the 
potential landscapes as well as on dissipation.

The techniques developed in this paper  
may be of particular interest for
super-heavy systems in which macroscopic theories are presently widely 
used\cite{Mol97,Ari99,Ada99,Smo99}. 
There, friction coefficients are often calculated assuming 
thermal equilibrium of other degrees of freedom. Such an assumption is not 
included in TDHF and it would be desirable to compare 
the two approach.

{\bf Acknowledgments}
      
The author thank Paul Bonche for helpful discussion at the initial 
stage of this work and for providing the three dimensional TDHF 
code. Usefull discussion with C\'edric Simenel, St\'ephane Gr\'evy 
and Olivier Juillet are acknowledged. We finally thank 
Dominique Durand for a carefull reading of the manuscript.
 
\appendix

\section{Inversion of equation with $\Delta R_{np}$}

Using three TDHF trajectories, 
noted ( {\footnotesize I}, {\footnotesize II} and
{\footnotesize III}), Eq. \ref{clas_np} can be interpreted
as a matricial equation which reads for a given position $R$: 
\begin{eqnarray}
\left(
\begin{array}{ccc}
1 &  -\frac{d R^{I}}{dt}  &
  -\frac{d \Delta R^{I}_{np}}{dt} \\
  & & \\
1 &  -\frac{d R^{II}}{dt}  &
 -\frac{d \Delta R^{II}_{np}}{dt} \\
  & & \\
1 &  -\frac{d R^{III}}{dt} & 
-\frac{d \Delta R^{III}_{np}}{dt} \\
\end{array}
\right) 
\left(
\begin{array}{c}
F\left( R \right) \\
\\
\gamma (R) \\
\\
\gamma_{np} (R) \\
\end{array} 
\right)
=
\left(
\begin{array}{c}
 \frac{dP^{I}}{dt} \\
 \\
 \frac{dP^{II}}{dt} \\
 \\
 \frac{dP^{III}}{dt}  \\
\end{array} 
\right)
\label{inver}
\end{eqnarray}
which we write ${\cal M} {\cal X} = {\cal B}$ in the following. The inversion 
of \ref{inver} is made difficult due to the fact that the additional
degree of freedom acts intermittently on the other 
components. When the coupling between $\Delta R_{np}$ of 
freedom is small, the set of equations becomes overcomplete.

\begin{figure}
\begin{center}
\epsfig{file=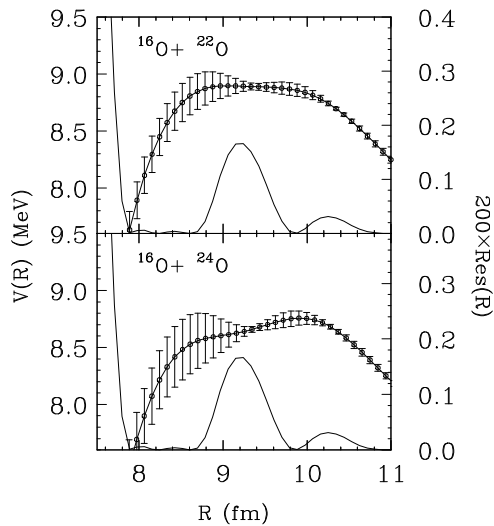,height=7.cm}
\end{center}
\caption{Example of calculated residue for the reactions 
$^{16}$O+$^{22}$O (top), $^{16}$O+$^{24}$O  (bottom) 
as a function of the relative distance. 
The potential obtained by inversion of equation \ref{eq:inv} 
is also superimposed with errorbars.}
\label{fig:residu}
\end{figure}

In order to solve this difficulty, we first consider two TDHF trajectories
and solve the two-dimensional Eq.  \ref{eq:inv}. 
This gives an approximative solution of \ref{inver} 
$\left( F_0(R) , \gamma_0 (R)\right)$. A measure of the
importance of $\Delta R_{np}$ is then obtained by computing the 
residue of the three dimensional equation:
\begin{eqnarray}
Res(R) = \left|\left| {\cal M} {\cal X}_0 - {\cal B} \right| \right|^2 
\end{eqnarray}  
where ${\cal X}_0 =  ^{t}( F_0(R) , \gamma_0 (R), 0)$. 
Example of calculated residue for the  
$^{16}$O+$^{22}$O and $^{16}$O+$^{24}$O reactions are presented in
Fig. \ref{fig:residu}. 
We see that a large residu at a given time (or equivalently
position) initiate the errors for future time (or lower position). This 
comes from the fact that we compute $V(R)$ by integration of $F$ over R
from large to small R values.

In this figure, the appearance of larger and larger residues during 
the approaching phase towards the fused systems is clearly
observed. This effect may be seen as the
starting of neutron-proton oscillations. We also remark that in between the
maxima, the residue is very close to zero and equation \ref{inver} reduces 
to Eq. \ref{eq:inv}. 
In our numerical implementation, when the residue is large, 
we solve equation \ref{inver} while for very small values, 
we switch to Eq. \ref{eq:inv}.

\end{document}